\documentclass[cits]{PoS}
\usepackage{amsmath} 
\usepackage{bbm} 
\allowdisplaybreaks 

\newcommand{\la}[1]{\label{#1}}

\newcommand{\eq}{Eq.~}
\newcommand{\se}{Sec.~}

\newcommand{\eqs}{Eqs.~}
\newcommand{\nr}[1]{(\ref{#1})}
\newcommand{\nn}{\nonumber \\}
\renewcommand{\(}{\left(}
\renewcommand{\)}{\right)}
\newcommand{\lb}{\left\{}
\newcommand{\rb}{\right\}}
\newcommand{\lk}{\left[}
\newcommand{\rk}{\right]}
\newcommand{\e}{\epsilon}
\newcommand{\sumint}[1]{\hbox{$\sum$}\!\!\!\!\!\!\!\!\int_{#1}}
\newcommand{\sumintp}[1]{\hbox{$\sum$}\!\!\!\!\!\!\!\!\int_{#1}{}^{\!\!\!\prime}}
\newcommand{\sumintz}[1]{\hbox{$\sum$}\!\!\!\!\!\!\!\!\int_{#1}\delta_{#1_0}}

\newcommand{\gammaE}{{\gamma_{\mathrm E}}}
\newcommand{\CA}{C_{\mathrm{A}}}
\newcommand{\order}[1]{{\cal O}(#1)}
\renewcommand{\vec}[1]{{\mathbf{#1}}}
\newcommand{\msump}[1]{\sum_{#1}{}^{\!\prime}}
\newcommand{\intp}[1]{\int\!\!\frac{\mathrm d^3 \vec{#1}}{(2\pi)^3}}
\newcommand{\intr}[1]{\int\!\!\mathrm d^3 \vec{#1}}
\newcommand{\p}[2]{\Pi_{#1}^{#2}}
\newcommand{\Ei}{\mathrm{Ei}}
\newcommand{\plog}[2]{\,\mathrm{Li}_{#1}\(#2\)}
\newcommand{\intx}[1]{\int_0^{\infty}\!\!\!\!\!\!\mathrm d #1}
\newcommand{\fact}[1]{\frac{#1}{(4 \pi)^6}}

\newcommand{\Ka}{J_{220011}^{000}} 
\newcommand{\Kb}{J_{310011}^{000}} 
\newcommand{\Kc}{J_{320011}^{002}} 
\newcommand{\Kd}{J_{410011}^{020}}
\newcommand{\Ke}{J_{510011}^{220}} 
\newcommand{\Kf}{J_{510011}^{400}} 
\newcommand{\Kg}{J_{710011}^{800}} 
\newcommand{\Kh}{J_{730-111}^{730}} 
\newcommand{\Ki}{J_{820-111}^{820}} 
\newcommand{\Va}{J_{121110}^{000}}
\newcommand{\Vb}{J_{211110}^{000}}
\newcommand{\Vc}{J_{221110}^{002}}
\newcommand{\Vd}{J_{311110}^{020}}
\newcommand{\Ve}{J_{411110}^{022}}
\newcommand{\Idaa}{I_4 I_1 I_1}
\newcommand{\Icba}{I_3 I_2 I_1}
\newcommand{\Ibbb}{I_2 I_2 I_2}

\title{Automated computation meets hot QCD}

\ShortTitle{Automated computation meets hot QCD}

\author{Ioan Ghi\c{s}oiu, \speaker{York Schr\"oder}\\
Faculty of Physics, University of Bielefeld, 
33501 Bielefeld, Germany\\
E-mail: \email{yorks@physik.uni-bielefeld.de}}

\abstract{We give a short review on recent progress in the field of 
automated calculations in finite-temperature field theory, where
integration-by-parts techniques have proven (almost) as useful as in the 
zero-temperature case.
Furthermore, we provide one concrete example of an evaluation of a
new three-loop master sum-integral that exhibits maximal divergence.}

\FullConference{Loops and Legs in Quantum Field Theory - 11th DESY 
Workshop on Elementary Particle Physics\\
April 15-20, 2012\\
Wernigerode, Germany}

\begin{document}


\section{Introduction}

In ordinary zero-temperature perturbative field theory,
this conference series has witnessed a rapid development
of automated computer-algebra aided approaches over the past
decade or so. 
Triggered perhaps by the integration-by-parts (IBP) 
method \cite{Chetyrkin:1981qh,Tkachov:1981wb}
that received an enormous boost from its algorithmic description
by Laporta \cite{Laporta:2001dd}, the problem of reducing the large variety
of Feynman integrals that occur in a given perturbative computation
to a small number of master integrals can nowadays be regarded as solved,
in the sense that there exist independent public computer-algebraic
implementations \cite{Anastasiou:2004vj,Smirnov:2008iw,vonManteuffel:2012yz} which, 
given sufficient hardware resources, 
perform this step in an automated way.
Concerning the subsequent evaluation of master integrals, the 
degree of automation is less developed -- after all, integration
is still an art. However, besides a much better understanding of
the analytic structure of the numbers and functions that are
contained in the perturbative series \cite{Remiddi:1999ew,Vermaseren:1998uu,Moch:2001zr,Ablinger:2011te}, 
there has been progress in developing powerful approaches to
obtain analytic results \cite{Laporta:2001dd,Argeri:2007up}
as well as stable and fast numerical tools
\cite{Smirnov:2009pb,Borowka:2012yc}
that promise to be applicable to large classes of Feynman integrals.

In this note, we will discuss the corresponding situation
for finite-temperature perturbation theory.
The present state of affairs is that, while many of the methods 
and algorithmic tools developed for zero-temperature field theory
(such as generation and classification of Feynman 
graphs,
efficient color and Lorentz algebra 
as well as IBP reduction methods) 
can be -- and have been -- applied with only minor adjustments to
finite-temperature systems, only very few master sum-integrals
that remain after the IBP reduction step are known 
(see, e.g.\ Refs.~\cite{Arnold:1994ps,Gynther:2007bw,Gynther:2009qf,Andersen:2008bz,Moller:2010xw} 
or the review \cite{Schroder:2012hm}).
It seems difficult to profit from the comparably mature zero-temperature
techniques for this step, due to the very different analytical
structure that the sums bring about. Here, we will therefore give
only a brief discussion on thermal IBP methods, to concentrate 
then on sum-integral evaluation, producing a new result for 
a phenomenologically relevant case.


\section{Sum-integral reduction via IBP}

The basic principle of the IBP method \cite{Chetyrkin:1981qh,Tkachov:1981wb} 
applies to any integral, 
and can hence also be applied to the sum-integrals that occur 
in finite-temperature field theories. The Matsubara sums
are simply left untouched, while their summands are interpreted
as massive loop integrals in a reduced space-time dimension and
with masses provided by the Matsubara frequencies, with the IBP
relations acting upon these massive loop integrals.
A more detailed review of these techniques is given e.g.\ in 
Ref.~\cite{Nishimura:2012ee}. 
Similar to the zero-temperature case, the linear relations
among sum-integrals -- as generated by IBP on the Matsubara summands --
can be systematically used for a sum-integral reduction step,
employing a variant of the Laporta algorithm \cite{Laporta:2001dd}.
This approach has been successfully used for a number of higher-order
calculations in finite-temperature QCD (for examples, 
see e.g.\ \cite{Nishimura:2012ee,Laine:2005ai,Moeller:2012da}), 
to enable basis transforms among master sum-integrals (see below),
or even to aid in managing the infrared (IR) behavior of
Matsubara summands when evaluating master integrals \cite{Ghisoiu:2012kn}.
To show two concrete examples, IBP has revealed that a non-trivial
2-loop sunset-sum-integral \cite{Schroder:2008ex} 
and a non-trivial 3-loop mercedes-sum-integral
vanish identically:
\begin{align}
\sumint{PQ}\frac{1}{P^2\,Q^2\,(P-Q)^2}\stackrel{\rm IBP}{=}0\;,\qquad
J_{111111}^{000}\stackrel{\rm IBP}{=}0\;,
\end{align}
where we have used a generic notation for massless bosonic 3-loop vacuum 
sum-integrals
\begin{align}
\la{eq:Jdef}
J_{abcdef}^{\alpha\beta\gamma} &\equiv 
\sumint{PQR} 
\frac{(P_0)^{\alpha} (Q_0)^{\beta} (R_0)^{\gamma}}
{[P^2]^a [Q^2]^b [R^2]^c [(P-Q)^2]^d [(P-R)^2]^e [(Q-R)^2]^f}\;.
\end{align}
In our notation, bosonic (Euclidean) 
four-momenta are denoted by $P=(P_0,\vec{p})=(2\pi n T,\vec{p})$, 
with $T$ being temperature of the thermal system,
inverse propagators are $P^2 = P_0^2 + \vec{p}^{2}$, 
and the sum-integral symbol stands for
\begin{align}
\sumint{P} \equiv
T\sum_{n\in\mathbbm{Z}}\int\!\!\frac{\mathrm{d}^d\vec p}{(2\pi)^d}\;,
\quad\mbox{with~~}d=3-2\e \;.
\end{align}
A class of 1-loop bosonic tadpoles that we will need below 
can be evaluated analytically as
\begin{align}
\la{eq:I}
I_s \equiv \sumint{Q} \frac{1}{[Q^2]^s} 
= \frac{2T\,\zeta(2s-d)}{(2\pi T)^{2s-d}}\,
\frac{\Gamma(s-\frac{d}2)}{(4\pi)^{d/2}\Gamma(s)}\;.
\end{align}

In Ref.~\cite{Moeller:2012da}, it has been shown that the
computation of NNLO corrections to the spatial string tension
of pure Yang-Mills theory
(mapped to Taylor coefficients of background-gauge-field 
self-energies \cite{Laine:2005ai})
can be reduced to 3-loop basketball-type master
sum-integrals $K_i$ (and products of simpler 1-loop ones $I_i$),
giving the gauge-invariant expression
\begin{align}
\Pi_{\rm T3}'(0) &=
\CA^3\(\sum_{i=1}^9 \beta_i(d)\,K_i
+\beta_{10}(d)\,\Idaa+\beta_{11}(d)\,\Icba+\beta_{12}(d)\,\Ibbb\)
\\
\la{eq:Klist}
\{K_1,\dots,K_9\} &\equiv
\{\Ka,\Kb,\Kc,\Kd,\Ke,\Kf,\Kg,\Kh,\Ki\}\;,
\end{align}
where the coefficients $\beta_i$ are rational functions that contain
simple and double poles in $1/(d-3)$, and where all 
massless bosonic 3-loop vacuum sum-integrals $K_i$ have been represented 
in terms of the notation introduced in \eq\nr{eq:Jdef}.
After an extensive reverse search in our IBP database, it proves possible to
transform the above expression into the equivalent representation
\begin{align}
\la{eq:newBasis}
\Pi_{\rm T3}'(0) &=
\CA^3\(\sum_{i=1}^6 r_i(d)\,V_i+r_7(d)\,\Idaa+r_8(d)\,\Icba+r_9(d)\,\Ibbb\)
\\
\la{eq:Vlist}
\{V_1,\dots,V_6\} &\equiv
\{\Va,\Vb,\Vc,\Vd,\Ve,\Kb\}
\end{align}
which, much in the spirit of the epsilon-finite basis advocated in
Ref.~\cite{Chetyrkin:2006dh}, does not contain divergences in the coefficients
$r_i(d)$ as $d\to3$ (note that $V_6=K_2$ was already contained in the 
old basis listed in \eq\nr{eq:Klist}, and was kept since it is has 
already been evaluated in Ref.~\cite{Moller:2010xw}).
\eq\nr{eq:newBasis} can now serve as a convenient starting point for
determining the spatial string tension, once the five unknown master
sum-integrals $\{V_1,\dots,V_5\}$ have been evaluated up to their 
constant parts. For $V_1$, see the next section.


\section{A 3-loop master sum-integral of dimension zero}

Let us now turn to a concrete example of 3-loop sum-integrals,
to demonstrate the main techniques that are used in the master 
integral evaluation step. Here, we work in $d=3-2\e$ spatial 
dimensions, and wish to evaluate the $\e$\/-expansion of $V_1$ 
given in \eq\nr{eq:Vlist} up to the constant term.

$V_1$ is a sum-integral of spectacles-type, for which a generic evaluation
procedure has been discussed in Ref.~\cite{Ghisoiu:2012yk}.
Subtracting subdivergences,
the sum-integral is split into its 
finite part (which can typically be computed only 
numerically in configuration space),
and its 
divergent part (which can be expressed analytically in terms of 
Zeta and Gamma functions),
treating the $P_0=0$ mode separately (softening the IR via
IBP relations if needed).
We thus decompose $V_1$ as
\begin{align}
V_1 &\equiv J_{12111}^{000} = 
\sumint{P} \frac1{P^2}\,\p{21}{}(P)\,\p{11}{}(P) \\
\la{eq:v1split}
&= \sumintp{P} \frac1{P^2}\Big\{
(\p{21}{} -\p{21}{E})(\p{11}{}-\p{11}{B}) 
+\p{21}{E}(\p{11}{}-\p{11}{B}) 
+(\p{21}{} -\p{21}{E}-\p{21}{C})(\p{11}{B} -\p{11}{D})\Big\}\,+\nn
&+\sumintp{P} \frac1{P^2}\Big\{
(\p{21}{} -\p{21}{E})\p{11}{D}
+\p{21}{C}(\p{11}{B}-\p{11}{D}) 
+\p{21}{E} \p{11}{B}\Big\}
+\sumintz{P} \frac{\p{21}{} \p{11}{}}{P^2} \;,
\end{align}
where the first line collects the finite pieces (the primed sum excludes the
$P_0=0$ term), the second line the divergent pieces as well as the zero-mode, 
and we have used the 1-loop 2-point structures
\begin{align}
\la{eq:Pi1}
&\p{ab}{}(P)\equiv\p{ab}{}\equiv\sumint{Q}\frac1{[Q^2]^a[(P-Q)^2]^b}
\;,\quad
\p{ab}{B}\equiv\int_Q\frac1{[Q^2]^a[(P-Q)^2]^b}=
  \frac{G(a,b,d+1)}{[P^2]^{a+b-(d+1)/2}}
\;,\\
&\mbox{where~~}G(s_1,s_2,d) =
\frac{\Gamma(\frac{d}2-s_1)\Gamma(\frac{d}2-s_2)\Gamma(s_1+s_2-\frac{d}2)}
{(4\pi)^{d/2}\Gamma(s_1)\Gamma(s_2)\Gamma(d-s_1-s_2)} \;,\\
\la{eq:Pi2}
&\p{ab}{C}\equiv\p{ab}{B}+\frac{I_a}{[P^2]^b}+\frac{I_b}{[P^2]^a}
\;,\quad
\p{ab}{D}\equiv\frac{[P^2]^\e}{(2\pi T)^{2\e}}\,\p{ab}{B}
\;,\quad
\p{ab}{E}\equiv\sumintz{Q}\frac1{[Q^2]^a[(P-Q)^2]^b}
\;.
\end{align}
We will in the following evaluate the various contributions 
to $V_1$ of \eq\nr{eq:v1split} in turn.


\subsection{Finite pieces}
\la{se:fin}

Using inverse 3d Fourier transforms of the 2-point functions, the
first term of \eq\nr{eq:v1split} reads
\begin{align}
& \sumintp{P} \frac1{P^2}
\times(\p{21}{} -\p{21}{E})
\times(\p{11}{}-\p{11}{B}) \nn
&= T \msump{P_0} \!\intp{p} \frac{1}{P^2} 
\!\times\!\frac{T}{2(4 \pi)^2} \!\intr{r} \frac{e^{i \vec{pr}}}{\bar{r}}  
\msump{Q_0} \frac{e^{-\!|Q_0|r\!-\!|Q_0\!+\!P_0|r}}{|\bar Q_0|}
\!\times\!\frac{T^3}{4} \!\intr{s} \frac{e^{i \vec{ps}}}{\bar s^2} 
\Big(\!\!\coth \bar s \!-\!\frac{1}{\bar s}\Big) e^{-|P_0|s}
+\order{\e}\nn
&=\fact{2} \msump{n,m} \intx{x} \intx{y}\,\frac{1}{y} 
\( \coth y -\frac{1}{y} \) \frac{e^{-(|n|+|m|+|m+n|)y}}{|n| |m|}
\( e^{-|n||x-y|}-e^{-|n|(x+y)}\)
+\order{\e}\nn
&= \fact{1} \intx{y}\,\frac1{3y}\,\( \coth y -\frac{1}{y} \)
\Big[
6y\big[\ln(1-e^{-2y})+y\big]^2
-\pi^2\big[\ln(1-e^{-2y})+4y\big]
-14y^3
\,+\nn&+
6\ln(1\!-\!e^{-2y})\plog{2}{e^{-2y}}
+12y\big[\plog{2}{1/(1\!-\!e^{-2y})}-i\pi\ln(1\!-\!e^{-2y})\big]
-6\plog{3}{1-e^{2y}}\Big]+\order{\e}
\nn
\la{eq:fin1}
&= \fact{c_1} +\order{\e}\,,\quad c_1\approx 0.6864720593640618954(1)\;,
\end{align}
where we have used dimensionless variables such as 
$\bar r=2\pi T|\vec r|=x$ and $\bar Q_0=Q_0/(2\pi T)=m$;
performed the momentum integration as well as the 
angular integration in configuration space in the third line;
integrated over $x$ and performed both sums in the fourth line;
and performed the last step simply
via Mathematica's numerical integration routine.

In full analogy\footnote{Here, 
we use $\p{21}{E}\stackrel{\rm IBP}{=}
(d-3)\frac{2P_0^2-P^2}{[P^2]^2}\,\p{11}{E}
-\frac{2\,P_0^2}{[P^2]^2}\,\p{02}{E}
\stackrel{d\!=\!3}{=}-\frac{T\,|P_0|}{4\pi[P^2]^2}$, 
where we have treated $|P_0|$ as a mass in the 3d 1-loop self-energy integrals
$\p{ab}{E}$, and noticed that $\p{11}{E}$ is finite at $d=3$,
and that the tadpole $\p{02}{E}$ is known analytically.},
we get for the second term of \eq\nr{eq:v1split}
\begin{align}
& \sumintp{P} \frac1{P^2}
\times\p{21}{E}
\times(\p{11}{}-\p{11}{B})\nn
&=T\msump{P_0} \intp{p} \frac{1}{P^2} 
\times\( -\frac{T}{4\pi} \frac{|P_0|}{P^4} \) 
\times\frac{T^3}{4} \intr{r} \frac{e^{i \vec{pr}}}{\bar r^2}  
\( \coth \bar r -\frac{1}{\bar r} \)e^{-|P_0|r}+\order{\e}\nn
&=-\fact{2} \sum_{n=1}^\infty \intx{x} \( \frac{x}{n} +\frac{1}{n^2}\) 
\( \coth x -\frac{1}{x} \) e^{-2nx}+\order{\e}\nn
\la{eq:fin2}
&=\fact{1} \Big[ 
\frac{\pi^2}{6} 
+\frac{3}{2} \zeta(3)
+\frac{\pi^2}{3} \ln 2\pi 
-4\pi^2 \ln G \Big]+\order{\e}\;,
\end{align}
where $G\approx 1.28243$ is the Glaisher constant.
Similarly, the third term of \eq\nr{eq:v1split} evaluates as
\begin{align}
& \sumintp{P} \frac1{P^2}
\times(\p{21}{} -\p{21}{E}-\p{21}{C})
\times(\p{11}{B} -\p{11}{D})\nn
&=T\msump{P_0} \intp{p} \frac{1}{P^2}
\times\frac{T}{2(4\pi)^2} \intr{r} \frac{e^{i \vec{pr}}}{\bar r} 
\Big[ f_{210}(\bar r,|\bar P_0|)\Big]
\times\( -\frac{\ln \bar P^2}{8 \pi^2} \) 
+\order{\e}\nn
&\mbox{~~~~~~~with~~}
f_{210}(x,n) \equiv e^{2nx} B(e^{-2x},n\!+\!1,0) 
+ H_{n}-\ln(1\!-\!e^{-2x}) 
+e^{2nx} \Ei(-2nx) 
+\ln\frac{2x}{n}-\gammaE  -\frac{x}{6 n}\nn
\la{eq:fin3int}
&=\fact{8} \sum_{n=1}^\infty \intx{x}\,e^{-2nx}
\lk e^{2nx}\Ei(-2nx)+\ln\frac{x}{2n}+\gammaE \rk 
\Big[ f_{210}(x,n)\Big]
+\order{\e}\\
\la{eq:fin3}
&= \fact{c_2} +\order{\e}\;,\quad c_2\approx-6.404(1)\;,
\end{align}
where we have used the incomplete Beta function 
$B(z,n+1,0)=\int_0^z\!{\rm d}t\,t^n/(1-t)=
-\ln(1-z)-\sum_{m=1}^{n}z^m/m$, 
harmonic numbers $H_n = \sum_{m=1}^{n} 1/m$
and exponential integral
$\Ei(z)=\int_{-\infty}^z\!{\rm d}t\,e^t/t$.
After integrating over $\vec p$ the angular $\vec r$ integration 
was trivial, leaving \eq\nr{eq:fin3int}
where the summation converges somewhat slowly 
and the evaluation of the integrand itself is costly since it contains 
special functions; for the numerical precision given above,
we have truncated the sum at $n_{\rm max}=7000$
and estimated the remainder by fitting to a power-law $a/n^b$
in the interval $n\in[7000,19000]$ (obtaining $b\approx1.9$) 
and summing this fit to infinity.


\subsection{Divergent pieces}
\la{se:div}

Introducing the following 2-loop vacuum sum-integrals
(see e.g.\ Appendix A and B of Ref.~\cite{Ghisoiu:2012yk})
\begin{align}
A(s_1,s_2,s_3;s_4;d) &\equiv \sumint{PQ} 
\frac{\delta_{Q_0}|P_0|^{s_4}}{[Q^2]^{s_1}[P^2]^{s_2}[(P-Q)^2]^{s_3}}\\
&=
\frac{2T^2\,\zeta(2s_{123}-2d-s_4)}{(2\pi T)^{2s_{123}-2d-s_4}}\,
\frac{\Gamma(s_{13}-\frac{d}2)\Gamma(s_{12}-\frac{d}2)
\Gamma(\frac{d}2-s_1)\Gamma(s_{123}-d)}
{(4\pi)^d\Gamma(s_2)\Gamma(s_3)\Gamma(d/2)\Gamma(s_{1123}-d)}
\;,\nn
L(s_1,s_2,s_3;s_4,s_5;d) &\equiv
\sumint{PQ} \frac{(P_0)^{s_4}\;(Q_0)^{s_5}}{[P^2]^{s_1}[Q^2]^{s_2}
[(P+Q)^2]^{s_3}}\;,\\
L(211;00;d) &\stackrel{\rm IBP}{=} -\frac{1}{(d-5)(d-2)}\,I_2\,I_2\;,
\end{align}
and noting from \eqs\nr{eq:Pi1},\nr{eq:Pi2} that
$\p{}{D}\sim\p{}{B}\sim1/[P^2]^x$ are simple powers, 
the fourth, fifth and sixth
terms of \eq\nr{eq:v1split} are effectively 2-loop structures
and can be evaluated analytically as
\begin{align}
\la{eq:div1}
&\sumintp{P} \frac1{P^2}\,(\p{21}{}-\p{21}{E})\,\p{11}{D} 
=\frac{G(1,1,d+1)}{(2\pi T)^{3-d}} 
\Big[ L(211;00;d)-A(121;0;d)-A(211;0;d)\Big]\;, \\
\la{eq:div2}
&\sumintp{P} \frac1{P^2}\,\p{21}{C}\,(\p{11}{B}-\p{11}{D})
= G(1,1,d+1) \Big[ G(2,1,d+1) \hat I_{5-d} 
+I_2 \hat I_{(7-d)/2} 
+I_1 \hat I_{(9-d)/2}\Big]\;, \\
\la{eq:div3}
&\sumintp{P}\frac1{P^2}\,\p{21}{E}\,\p{11}{B} 
=G(1,1,d+1)\,A(2,(5-d)/2,1;0;d)\;,
\end{align}
where we have used the abbreviation 
$\hat I_s\equiv I_s-(2\pi T)^{d-3}I_{s+(d-3)/2}$.
All three expressions contain poles as $d\rightarrow3$; 
the first two contribute starting from $1/\e^3$, 
the last one merely from $1/\e$. 


\subsection{Zero mode}

The seventh term of \eq\nr{eq:v1split} can be decomposed 
into finite (first line) and divergent parts as
\begin{align}
\la{eq:zeroMode}
\sumintz{P} \frac1{P^2}\,\p{21}{}\,\p{11}{} 
&= \sumintz{P} \frac1{P^2}\,(\p{21}{} -\p{21}{E})\,
(\p{11}{} -\p{11}{B} -\p{11}{E})
\,+\nn
&+\sumintz{P} \frac1{P^2} \Big\{ \p{21}{E}\p{11}{} 
 +\p{21}{}\,(\p{11}{B} +\p{11}{E})
-\p{21}{E}\,(\p{11}{B} +\p{11}{E})\Big\}\;.
\end{align}
In full analogy to \se\ref{se:fin}, 
the first (finite) term is treated in 3d coordinate space, as
\begin{align}
&\sumintz{P} \frac1{P^2}
\times(\p{21}{} -\p{21}{E})
\times(\p{11}{} -\p{11}{B} -\p{11}{E})\nn
&= T \intp{p} \frac{1}{\vec p^2}
\times\frac{T}{2(4\pi)^2} \intr{r}\, \frac{e^{i \vec{pr}}}{\bar r} 
\lk -2 \ln(1-e^{-2 \bar r})\rk 
\times\frac{T^3}{4} \intr{s}\, \frac{e^{i \vec{ps}}}{\bar s ^2} 
\( \coth \bar s -\frac{1}{\bar s}-1\)
+\order{\e}\nn
&= -\fact{4} \intx{x} \intx{y}\; \frac{x+y-|x-y|}{y} \ln(1-e^{-2x})
\( \coth y -\frac{1}{y} -1\)
+\order{\e}\nn
&=\fact{1} \frac{4}{3} \intx{y}\, \frac{1}{y} \( \coth y -\frac{1}{y}-1\) 
\Big[4y^3 -2\pi^2 y +3\big[\plog{3}{e^{2y}}+2\pi i\,y^2\big] -3\zeta(3)\Big]
+\order{\e}\nn
\la{eq:zero1}
&= \fact{c_3}+\order{\e}\;,\quad c_3\approx 10.33244698246374834(1)\;.
\end{align}
In full analogy to \se\ref{se:div}, 
the second (divergent) part of \eq\nr{eq:zeroMode} 
contains 1- and 2-loop structures $G$ and $A$ only 
and is hence known fully analytically
in terms of Gamma and Zeta functions:
\begin{align}
&\sumintz{P} \frac1{P^2} \lb \p{21}{E}\p{11}{} 
 +\p{21}{}\,\p{11}{E} +\p{21}{}\,\p{11}{B} 
 -\p{21}{E}\,(\p{11}{B} +\p{11}{E})\rb\nn
\la{eq:zero2}
&= T\,G(2,1,d)\,A(4-d/2,1,1;0;d) 
 + T\,G(1,1,d)\,A(3-d/2,2,1;0;d)
\,+\nn
&+G(1,1,d+1)\,A((5-d)/2,2,1;0;d) 
-0_{\rm scalefree}\;.
\end{align}


\subsection{Result}

Collecting from \eqs\nr{eq:fin1},\nr{eq:fin2},\nr{eq:fin3},\nr{eq:div1},
\nr{eq:div2},\nr{eq:div3},\nr{eq:zero1},\nr{eq:zero2}
and expanding around $d=3-2\e$, we finally obtain
\begin{align}
V_1 &= \frac{1}{6(4\pi)^6} \(\frac{e^{\gammaE}}{4\pi T^2}\)^{3\e}
\bigg[ \frac1{\e^3}  +\frac3{\e^2} 
+\frac1\e \Big( 13 -6 \gammaE^2 + \frac{3 \pi^2}{4} -12 \gamma_1 
-3\zeta(3)\Big) \,+\nn
&+ \Big( 51 -42 \gammaE^2 
+4\pi^2 \Big( \frac{19}{16} +\ln(2\pi) -12 \ln G\Big) 
+2\ln(2)\big( 12 -12 \gammaE^2 -24 \gamma_1 -\zeta(3) \big) \,+\nn
&+ 6\gammaE \big(3\zeta(3)-4-4\gamma_1\big)  
-84 \gamma_1 -36 \gamma_2 +\frac{25}{2}\,\zeta(3) -16 \zeta'(3) 
+ 6 \big(c_1+c_2+c_3\big) \Big) 
+\order{\e} \bigg] \nn
&\approx \frac{1}{6(4\pi)^6} \(\frac{1}{T^2}\)^{3\e} 
\(\frac{1}{\e^3} -\frac{2.86143}{\e^2} + \frac{15.2646}{\e} 
+ 28.56(1) +\order{\e}\) \;,
\end{align}
with $(c_1+c_2+c_3) \approx 4.615(1)$,
where the numerical error is dominated by \eq\nr{eq:fin3}.
The analytic part of the result contains the Glaisher constant $G$, 
for which $12 \ln(G) = 1+\zeta'(-1)/\zeta(-1)$,
zeta values as well as the Stieltjes constants $\gamma_i$,
defined by $\zeta(1+\e)=1/\e+\gammaE-\gamma_1\e+\gamma_2\e^2/2+\order{\e^3}$.

Note that our 3-loop sum-integral $V_1$ 
does contain a $1/\e^3$ divergence, as would
be naturally expected from the fact that the 1-loop tadpoles of
\eq\nr{eq:I} diverge as $1/\e$ at most (recall that $d=3-2\e$ 
in our notation).
This is in fact the first example where we observe such 
behavior -- all previously known non-trivial 3-loop cases 
in fact started at order $1/\e^2$.


\section{Conclusions}

While automated methods for Feynman integral reduction work well
when applied to the sum-integrals that arise in
finite temperature perturbation theory, 
the step of evaluating the resulting master sum-integrals 
is in a much less mature state. 
Only a small number of non-trivial higher-loop 
sum-integrals are known so far, their evaluation resting on 
a case-by-case analysis with intricate subdivergence subtraction techniques,
such as demonstrated here on the example of a new bosonic 3-loop tadpole
that enters in a NNLO determination of the spatial string tension.
Typical results are partly analytic and partly numeric, since often
the constant terms of the Laurent series
cannot be obtained in closed form, but mapped
onto simple finite low-dimensional representations.

Future progress in the field of classification and evaluation
of sum-integrals would certainly give a boost to the field of
finite temperature field theory. It will be interesting to see
whether this progress originates again from a fruitful application of
zero-temperature methods (such as e.g.\ Mellin-Barnes representations, classes
of multiple nested sums and integrals such as harmonic polylogarithms,
or systematic numerical methods based on sector decomposition),
or whether completely new ideas and structures are needed.


The work of I.G.~is supported by the Deutsche
Forschungsgemeinschaft (DFG) under grant no.~GRK~881.
Y.S.~is supported by the Heisenberg program DFG, contract no.~SCHR~993/1. 



\end{document}